\documentclass[preprint2]{proto}
\usepackage{times}

\newcommand{\refs}{\par\noindent\hangindent=1pc\hangafter=1}
\def\ptp{peak-to-peak }

\def\ri{\mathrel{(R_{\mathrm{C}}-I_{\mathrm{C}})}}
\def\hamag{\mathrel{\mathrm{H}\alpha}}

\def\mag{\mathrel{\mathrm{mag}}}
\def\hain{\mathrel{\Delta (R_{\mathrm{C}}-\mathrm{H}\alpha)}}

\def\haindices{H$\alpha$-indices }

\def\msun{\mathrel{M_{\odot}}}

\begin{document}

\title{\textbf{\LARGE The Rotation of Young Low-Mass Stars and Brown Dwarfs}}

\author {\textbf{\large  William Herbst}}
\affil{\small\em Wesleyan University}

\author {\textbf{\large  Jochen Eisl\"offel}}
\affil{\small\em  Th\"uringer Landessternwarte, Tautenburg }

\author {\textbf{\large  Reinhard Mundt}}
\affil{\small\em Max Planck Institute for Astronomy, Heidelberg}

\author {\textbf{\large Alexander Scholz}}
\affil{\small\em  University of Toronto}

\begin{abstract}
\baselineskip = 11pt
\leftskip = 0.65in
\rightskip = 0.65in
\parindent=1pc

{\small 	We review the current state of our knowledge concerning the  
rotation and angular momentum evolution of young stellar objects and 
brown dwarfs from a primarily observational view point.  There has been a tremendous growth 
in the number of young, low-mass objects with measured rotation periods over the last five years, due to  the application of wide field imagers on 1-2 m class telescopes. Periods are typically accurate to 1\% and available for about 1700 stars and 30 brown dwarfs in young clusters. Discussion of angular momentum evolution also requires knowledge of stellar radii, which are poorly known for pre-main sequence stars. It is clear that rotation rates at a given age depend strongly on mass; higher mass stars (0.4-1.2 M$_\odot$) have longer periods than lower mass stars and brown dwarfs. On the other hand, specific angular momentum is approximately independent of mass for low mass pre-main sequence stars and young brown dwarfs. A spread of about a factor of 30 is seen at any given mass and age. The evolution of rotation of solar-like stars during the first 100 Myr is discussed. A broad, bimodal distribution exists at the earliest observable phases ($\sim$1 Myr) for stars more massive than 0.4 M$_\odot$. The rapid rotators (50-60\% of the sample) evolve to the ZAMS with little or no angular momentum loss. The slow rotators continue to lose substantial amounts of angular momentum for up to 5 Myr, creating the even broader bimodal distribution characteristic of 30-120 Myr old clusters. Accretion disk signatures are more prevalent among slowly rotating PMS stars, indicating a connection between accretion and rotation. Disks appear to influence rotation for, at most, $\sim$5 Myr, and considerably less than that for the majority of stars. This time interval is comparable to the {\it maximum} life time of accretion disks derived from near-infrared studies, and may be a useful upper limit to the time available for forming giant planets. If the dense clusters studied so far are an accurate guide, then the typical solar-like star may have only $\sim$1 Myr for this task.  There is less data available for very low mass stars and brown dwarfs but the indication is that the same mechanisms are influencing their rotation as for the solar-like stars. However, it appears that both disk interactions and stellar winds are less efficient at braking these objects.  We also review our knowledge of the various types of 
variability of these objects over as broad as possible a mass range with 
particular attention to magnetically induced cool spots and magnetically 
channeled variable mass accretion. \\~\\~\\~}

\end{abstract}

\section{\textbf{Introduction}}

The study of stellar rotation during the pre-main sequence (PMS) and
zero age main sequence (ZAMS) phase provides important clues to the solution
of the angular momentum problem of star formation. It is also intimately connected to the evolution of the circumstellar disk out of which planets are believed to form. The angular momentum problem, simply stated, is that the specific
angular momentum (j = J/M where J is angular momentum and M is mass) of
dense molecular cloud cores, the birth places of low-mass
stars, is 5-6 orders of magnitude higher than that of solar-type
stars on the ZAMS (e.g., {\it Bodenheimer}, 1995). We note that by the time they reach the age of the Sun ($\sim$5 Gyr), solar-type stars have lost an additional 1-2 orders of 
magnitude of j due to the cumulative effect of torques from their magnetized 
coronal winds. 

Undoubtedly, the full angular momentum problem is solved by a
combination of factors occurring throughout the star formation process, not just a single event at a specific time ({\it Bodenheimer}, 1995). 
Important processes at early stages include magnetic torques between the
collapsing molecular cloud core and the surrounding interstellar medium as well as 
the deposition of large amounts of angular momentum in the orbital motions of a circumstellar disk, planetary system and/or binary star. At later stages
the redistribution of angular momentum within the disk and {\it J}-loss 
by magnetically driven outflows and jets become important.
Recent HST/STIS studies of optical jets from T Tauri stars (TTSs) (e.g., {\it Bacciotti et al.}, 2002;
{\it Woitas et al.}, 2005) show that these jets are rotating. In the framework
of magnetically driven outflows, the derived rotation 
velocities of these jets imply large angular momentum loss rates.
The magneto-hydrodynamic (MHD) models suggested for the acceleration of bipolar outflows
from young stars also involve a strong magnetic coupling between
the stellar magnetic field and the inner parts of the circumstellar
accretion disk. Hence, it seems likely that the rotation of PMS stars will be influenced by these processes.

On the observational side, the first studies of rotation to include a substantial number of TTSs employed high resolution spectra to measure line broadening (e.g. {\it Vogel and Kuhi}, 1981; {\it Hartmann et al.}, 1986; {\it Bouvier et al.}, 1986). The resulting measurements of rotational velocity (V$_{rot}$) are inherently uncertain due to the unknown inclination of the stellar rotation axis to the line of sight. It is also difficult to measure V$_{rot}$ for slow rotators since there is very little line broadening in that case. Nonetheless, it was clear from these studies that, in general, TTSs rotated much slower (about a factor of 10, on average) than their critical velocities, although details of the distribution such as its true breadth and bimodal nature were not apparent. 

Starting in the mid-1980's the technique of directly measuring the rotation period from photometric monitoring of a spinning, spotted surface was applied to a growing number of TTSs ({\it Rydgren and Vrba}, 1983;  {\it Herbst et al.}, 1986, 1987;  {\it Bouvier and Bertout}, 1989). Through a proper analysis of these brightness
modulations, the rotation period (P), which is independent of inclination angle, can be measured to
an accuracy of about 1$\%$, even for the slowest rotators. We note that P, or its equivalent, angular velocity ($\omega = 2 \pi / {\rm P}$), is currently the most accurately known stellar parameter for most PMS stars and it is known for a large number of them.

The amount of available rotation period data has skyrocketed 
during the last 6 years. This advance occurred as a result of
the application of wide field optical imaging devices to the problem. For the first time, it became possible to simultaneously monitor hundreds of PMS stars in young open
clusters such as the Orion Nebular Cluster (ONC), NGC 2264 and IC 348 (see e.g., {\it Mandel and Herbst}, 1991;
{\it Attridge and Herbst}, 1992; {\it Choi and Herbst}, 1996; {\it Stassun et al.}, 1999; {\it Herbst et al.}, 2000a,2000b, 2001, 
2002; {\it Rebull}, 2001; {\it Carpenter et al.}, 2001; {\it Lamm et al.}, 2004, 2005; 
{\it Makidon et al.}, 2004; {\it Littlefair et al.}, 2005; {\it Kiziloglu et al.}, 2005; {\it Nordhagen et al.}, in preparation). 
These imaging devices also provided the first extensive
 measurements of rotation periods for very low-mass stars (VLMSs) and 
brown dwarfs (BDs) in PMS and ZAMS clusters ({\it Scholz}, 2004;  
{\it Scholz and Eisl\"offel}, 2004a,b, 2005).
Altogether $\sim$1700 periods of low-mass PMS stars and
young VLMSs and BDs are currently available in the literature. In addition, for several
ZAMS clusters there are extensive rotation period data available on
both solar-type stars (see references in {\it Herbst and Mundt}, 2005)
and VLMSs and BDs (see {\it Scholz}, 2004;  {\it Scholz and Eisl\"offel}, 2004b 
and references therein).

In this article we provide an overview
of the rotation period data currently available on PMS objects over 
the broadest possible mass range that can be investigated with the photometric monitoring technique, i.e. 
from about 1.5 M$_{\odot}$ down into the substellar mass regime.
Furthermore, we discuss how our present knowledge of the rotation
properties of these objects in ZAMS clusters constrains our
current understanding of their angular momentum evolution during
the first few Myr. Our chapter is structured as follows:
in Section 2 we will describe the observational
method employed in measuring rotation periods.
The next two sections  deal with the rotation properties and
the empirical results on angular momentum evolution for
solar-type stars, VLMSs and BDs. In Section 5 we review our knowledge of the variability of these objects, the importance of disk-rotation interactions and magnetically-channeled mass accretion
over as broad as possible a mass regime. Finally, in Section 6 we discuss the confrontation between observation and theory, in particular the model popularly known as ``disk-locking".

\bigskip

\section{\textbf{Observational Methods}}
\bigskip
The method of determining stellar rotation periods by monitoring the motion of a surface spot of very different temperature from the surrounding photosphere can be traced back to Christopher Scheiner and Galileo Galilei, who employed it to determine the rotation rate of the Sun at low latitude (see, for example, {\it Tassoul}, 2000). Modern application of this method exploits the photometric variability induced by the spot or spot group as it is carried around by the star's rotation. Sufficiently dense photometric monitoring over at least a couple of cycles will reveal periodicity in many PMS stars' brightness variations that can be linked with certainty to rotation ({\it Rydgren and Vrba}, 1983; {\it Bouvier et al.}, 1986; {\it Stassun et al.}, 1999; {\it Rhode et al.}, 2001). This method only works for stars of spectral class G, K or M ($\sim$1.5  M$_\odot$ and less) and for BDs. It is most effective for the mid-K to early-M stars ($\sim$0.5 M$_\odot$) where the spot amplitudes are largest (see Section 5). 
   
The value of j, the specific angular momentum at the surface, of a spherically symmetric, uniformly rotating star depends on only two variables -- rotation period (P) and radius (R). In general, we may expect a star's surface rotation rate to be a function of latitude, as is well known to be the case for the Sun. Remarkably, this effect has never been convincingly demonstrated for any T Tauri star (see, however, {\it Herbst et al.}, 2005). On the contrary, one finds that measured rotation periods are stable to within the errors of their determination, typically 1\% ({\it Cohen et al.}, 2004;  {\it Kiziloglu et al.}, 2005; {\it Nordhagen et al.}, in preparation). This presumably means that spots on PMS stars are generally confined to a small range of latitudes or that their surfaces are rotating in much more rigid fashion than the Sun, or both. It is normally assumed by astronomers working in this area that the measured periodicity in brightness is the rotation rate applicable to all latitudes on the surface of the star.

\begin{figure}[t]
\resizebox{\hsize}{!}
{\includegraphics {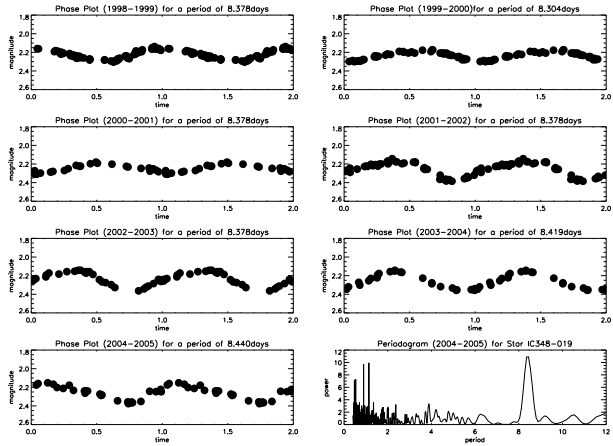}}
\caption{Light curve of HMW 19, a WTTS in IC 348, based on seven years of photometric monitoring in the Cousins I band at Wesleyan University ({\it Nordhagen et al.}, in preparation). The bottom right panel shows the periodogram function for one season. Note how the light curve shape and amplitude changes slightly from year to year while the period remains essentially constant to within the error of measurement.}
\label{star019}
\end{figure}

An example of the data is shown in Fig. \ref{star019} from the work of {\it Nordhagen et al.} (in preparation). Typically, one searches for periodicity using the Lomb-Scargle periodogram technique which is effective for unevenly spaced data sets. Evaluation of the {\it false alarm probability} (FAP) associated with a peak of any given power can be tricky and is best done with a Monte Carlo simulation of the data. Most authors in this field have adopted relatively conservative criteria generally equivalent to a FAP of about 0.01. Data obtained from a single observatory (longitude) have an unavoidable 1 d$^{-1}$ natural frequency embedded in them, imposed by the rotation of the Earth. Truly periodic objects therefore normally have more than one significant peak in their periodograms due to the beat phenomenon. Separating true periods from beat periods can be difficult and often requires qualitative judgments about which period does the best job of phasing the data. Most disagreements about periods in the literature arise from this complication. Continued monitoring, or a {\it v} sin {\it i} measurement will permit resolution of the question. Occasionally a star will show periodicity at one-half its true rotation period due to the existence of spots in opposite hemispheres of longitude. Examples are V410 Tau ({\it Vrba et al.}, 1988) and CB 34V ({\it Tackett et al.}, 2003).

It is expected on general physical grounds (e.g., {\it Tassoul}, 2000) that rotation rate will vary with depth into a star and helioseismology has shown this to be the case for the Sun, especially within its convective zone (e.g., {\it Thompson et al.}, 2003). Since we are very far from having any way of determining, either observationally or theoretically, what the dependence of rotation with depth in a PMS star might be like, there is realistically at present no way of measuring how j might vary in a radial sense. From a purely empirical point of view one can, therefore, discuss only the {\it surface value} of j. Models, of course, can be and have been constructed with assumed rotation laws (e.g., uniform with depth) and with interior mass distributions satisfying the usual constraints of stellar structure. A common assumption is that fully convective stars are rigid rotators, although rigid rotation is not what is observed in the solar convection zone. These assumptions, of course, do allow one to estimate a value of j applicable to the whole star at the expense of increased uncertainty due to the inability to test critical assumptions.

Finally, we note that the surface value of j for a spherical star (see {\it Herbst and Mundt}, 2005 for a discussion of non-spherical stars) depends on stellar radius, a notoriously difficult quantity to measure for PMS stars. Debates in the literature on rotational evolution during the PMS stage often center on how to evaluate the somewhat bewildering data on luminosity and effective temperatures of such stars, from which their stellar radii are inferred. Rotation periods are relatively easy to determine, are highly accurate, and are generally not the source of disagreement about interpretation in this field. Radii, on the other hand, are hard to determine for any individual star, show a large scatter among stars of apparently the same mass, within a single cluster, and are at the root of some recent debates in the literature over how to interpret data on stellar rotation ({\it Rebull et al.}, 2002, 2004; {\it Lamm et al.}, 2004, 2005; {\it Makidon et al.}, 2004; {\it Herbst and Mundt}, 2005). The problem is exacerbated by the fact that the relative ages of PMS stars of the same mass are set by their relative radii. If radii are in error then ages are in error and evolutionary trends become difficult to discern. We return to this difficulty in what follows but first give an an overview of recent empirical results on rotation.  
\bigskip

\section{\textbf{Rotation of Young Stellar Objects}}
\bigskip

\begin{figure}[t]
\resizebox{\hsize}{!}
{\includegraphics[angle=-90]{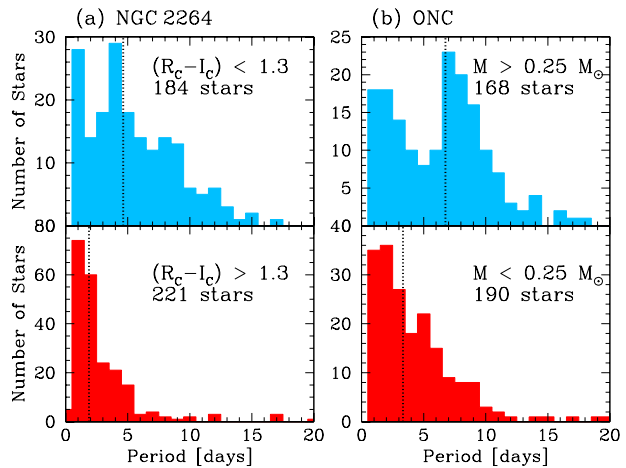}}
\caption{The rotation period distribution in NGC 2264 and the ONC divided by mass range. Vertical lines indicate the median value of each sample. It is clear that the higher mass stars in both clusters have longer rotation periods and exhibit bimodal distributions. It is also clear that, within both mass ranges, stars in the ONC tend to have longer rotation periods. The division for the ONC is actually by effective temperature and translates to 0.4 M$_\odot$ for some PMS models.}
\label{Lamm}
\end{figure}

With $\sim$1700 rotation periods measured, there is now a fairly good empirical understanding of spin rates of PMS stars and their dependence on spectral type (or mass). A couple of surprising results have emerged from this, the first of which is the breadth of the rotation period distribution, which extends
at least over a factor 30 for all well-observed mass ranges. The second surprising fact
is that the measured rotation period distributions are highly
mass-dependent. 
For stars with  0.4 $<$ M $<$ 1.5\  M$_{\odot}$, the period distribution ranges
from  $\sim$0.6\,d to  $\sim$20\,d and is clearly bimodal with peaks near 2 and 8 days in the ONC and near 1 and 4 days in NGC 2264 (see Fig. \ref{Lamm}). For 
stars with masses below 
0.4\,M$_{\odot}$ (or 0.25 \,M$_{\odot}$ depending on which PMS models one adopts) the median of the distribution is about a factor 2 shorter in both clusters ({\it Herbst et al.}, 2001; {\it Lamm et al.}, 2004, 2005).  

A recent comparison of 
the {\it j}-distributions of solar-type PMS stars in the ONC and NGC\,2264 
with solar-type ZAMS stars (0.4-1.2 M$_{\odot}$) shows that the 
{\it j}-distributions of the PMS and ZAMS sample
match very well for high {\it j}-values, {\it but not for low {\it j} values}
({\it Herbst and Mundt}, 2005). For the ZAMS stars the {\it j}-distribution extends
by about a factor 3 towards lower values.
This means that the rapid rotators among the optically observable
solar-like PMS stars
do not lose much, if any, angular momemtum on their way to the ZAMS, while
the slow rotators do continue to experience some braking. This again indicates that most of the angular momentum problem is
solved by the time these stars become observable in the optical.
There is some indication that much younger and much deeper embedded
PMS stars (i.e. protostars) have about a factor 2 higher
 V$_{rot}$ values than their optically visible counterparts
(see {\it Covey et al.}, 2005 and references therein) suggesting significant
angular momentum loss during this phase. Such an increased angular
momentum loss seems to be in accordance
with the substantial bipolar outflow activity during this phase.

As mentioned in the introduction, the measured periods vary by about a factor of 30 from 0.6 to about 20 days for solar-like stars. This wide range is present at all spectral types and masses and seen in all samples (clusters and associations) where enough data exist. A difficulty in interpreting these data is that PMS stars contract rather rapidly and if they conserve angular momentum will spin up rapidly. For example, a star with an 8 day rotation period at 1 Myr will have a 5 day rotation period at 2 Myr if it conserves angular momentum at its surface. It is, therefore, critical to compare samples at the same age, if one wants to discern a mass dependence. 

The dependence of rotation on mass among PMS stars of a common age was first demonstrated for the ONC by {\it Herbst et al.} (2001). They found that lower mass stars, in general, rotate faster than their higher mass counterparts. In terms of (surface) j, however, there is little or no dependence on mass. It appears that lower mass stars spin faster primarily because they have smaller radii. When separated at a spectral class of about M2, corresponding to a mass of between 0.25 and 0.4 solar masses (depending on the model adopted), one finds that the higher mass stars have a distinctly bimodal period distribution with peaks near 2 and 8 days, while the lower mass stars, in addition to having a shorter period median, may have a somewhat smoother distribution, perhaps characterized by a single mode, although this is uncertain.

{\it Lamm} (2003) and collaborators ({\it Lamm et al.}, 2004, 2005) found similar results for another young cluster in which the stars can reasonably be regarded as mostly coeval. His results are shown in Fig. 2, where the bimodal nature of the higher mass stars in both the ONC and NGC 2264 is clearly seen, as is the more rapid rotation and, perhaps, single mode distribution of the lower mass stars. The figure also demonstrates that, for PMS stars of the same mass range, those in NGC 2264 rotate about twice as fast as those in the ONC. Statistical analyses of the distributions confirms this claim; NGC 2264 stars cannot have been drawn from  the same parent population as ONC stars.

The interpretation of the difference in rotation rates between the clusters favored by Lamm and collaborators is that NGC 2264 is a factor of 2 older than the ONC and that a significant fraction of the stars have spun up as they contracted, roughly conserving angular momentum. At the same time, not all of the stars could have spun up since there remains a fairly well-populated tail of slow rotators at all masses among the NGC 2264 stars. There is no evidence that these slower rotators are, as a group, larger (i.e. younger) than the rapid rotators in the cluster, so an age spread within the cluster does not seem to account for the breadth of the rotation distribution. We discuss in more detail below, the interpretation of the broad, bimodal rotation distribution. First, however, we turn to a discussion of the evolution of the rotation of higher mass stars with time over a broader time frame ($\sim100$ Myr).

\bigskip

\section{\textbf{Angular Momentum Evolution: Empirical Results}}
\bigskip

Since the rotation rates of PMS stars are mass dependent it is only appropriate to discuss the evolution of j with time within restricted mass regimes. One such important regime, in part because it includes the Sun, is 0.4-1.2 M$_\odot$. Two other regimes are the very low mass PMS stars, corresponding to spectral class M2.5 and later, and the brown dwarfs. At present, the only mass range which is reasonably well constrained by observations is the solar-like range because ZAMS rotation periods are known for a significant number of such stars from studies of young clusters such as the Pleiades, IC 2602 and the $\alpha$ Per cluster (see references in {\it Herbst and Mundt}, 2005). Here we will discuss the time evolution of j for the 0.4-1.2 M$_\odot$ stars in some detail and then briefly say how this may depend on mass.

\bigskip

\noindent
\subsection{\textbf{Solar-like stars (0.4-1.2\,M$_{\odot}$)}}

\bigskip
\noindent
It has been known for more than a decade that ZAMS stars of around 1 M$_\odot$ display an enormous range of rotation rates, larger even than is seen among PMS stars. Photometric monitoring of spotted stars in three clusters with ages around 30-120 Myr has provided periods for about 150 ZAMS (or close to ZAMS) stars (see references in {\it Herbst and Mundt}, 2005). Radii are well known for these stars, so it is possible to determine the {\it j}-distribution with some certitude. This provides a ``goal" for the evolution of the PMS distributions.

Early studies of the evolution from PMS to MS assumed that stars started from a somewhat narrow {\it j}-distribution (e.g., {\it Bouvier}, 1994) and that the breadth observed on the ZAMS developed during the PMS phase. A common assumption was that all PMS stars had 8 day periods to start with, based on the larger peak in the distribution of the ONC stars and the corresponding peak for CTTS ({\it Bouvier et al.}, 1993). With increased data samples it is now clear that most of the breadth of the ZAMS population is already built in at the earliest observable PMS phases, represented by the ONC. The current problem divides into understanding how the broad PMS distribution at 1 Myr came into existence and how rotation evolves between the PMS and ZAMS. 

The first question is difficult to constrain with data because Class 0 and 1 proto-stars are rare and it is hard to ascertain essential data on them, in particular their masses, radii and rotation rates. No rotation periods have yet been discovered for proto-stars. Line broadening measurements, however, have been made for 38 Class I/flat spectrum objects by {\it Covey et al.} (2005).  Although the sample is necessarily small and heterogeneous, the authors do find an average {\it v} sin {\it i} for the sample of 38 km/s, which they argue is significantly larger than for CTTS. Since proto-stars should be, if anything, larger than CTTS, this implies that, as a group, they have larger values of surface angular momenta than CTTS. These data, therefore, reinforce the view that the protostellar or very early PMS stage is a time during which large amounts of angular momentum are lost. This is when the jets and winds which are needed to carry off angular momentum are most prominent and active. This is also the relatively brief period of time ($\sim$1 Myr) when a majority of stars must lose substantial amounts of angular momentum to reduce their j values to the levels observed in the ONC.

The second aspect of understanding the origin of the ZAMS {\it j}-distribution is more amenable to an observational approach, but is not without its own set of complications. The goal is to understand how a 1 Myr old {\it j}-distribution, as represented by the ONC, evolves to a 100 Myr old distribution, as represented by the ZAMS clusters and the problem is that there are very few signposts along the way. That is, one requires a substantial number of stars of homogeneous origin (or at least common age) and data on young clusters are scarce. At present, there are really only two PMS clusters with enough periods known to reasonably define the rotation distributions -- the ONC and NGC 2264. IC 348 may soon be included among these (see {\it Nordhagen et al.}, in preparation).

\begin{figure}[t]
\resizebox{\hsize}{!}
{\includegraphics{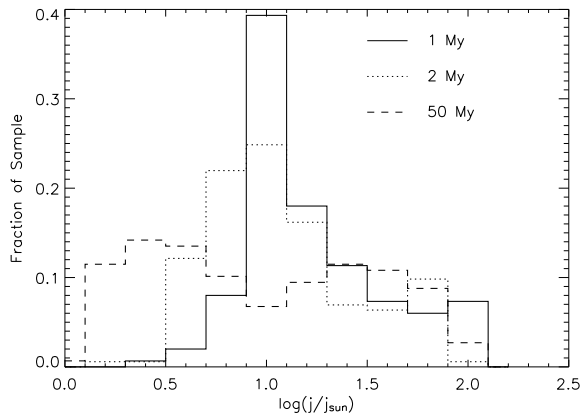}}
\caption{The observed
j-diNstribution of the ONC, NGC 2264, and combined three other
clusters corrected for wind losses.
It is clear that there is little change on the high-j side, implying
that rapidly rotating stars nearly conserve angular momentum as they
evolve from the PMS to the MS. However, there is a broadening of the
distribution on the low-j side which is already noticeable in the comparison
of the 1 My old ONC with the 2 My old NGC 2264 cluster and becomes
quite dramatic when comparing the PMS and ZAMS clusters. This indicates
that slowly rotating PMS stars must lose substantial additional
amounts (factor of 3 or more) of their surface angular momentum
during contraction to the ZAMS.}
\label{jhists}
\end{figure}

Figure \ref{jhists} shows the observed situation for these clusters and for the set of ZAMS (or nearly ZAMS) stars drawn from the Pleiades, IC 2602 and the $\alpha$ Per clusters ({\it Herbst and Mundt}, 2005). There are many {\it caveats} and considerations in comparing these samples and the interested reader is referred to the original work for the details. Here we touch only on the main points, the first of which is that there are clear, statistically significant differences between the three {\it j}-distributions represented by the ONC, NGC 2264 and the ZAMS clusters. If one assumes that each sample had similar initial rotation parameters, then these differences can be interpreted as an evolutionary sequence.

The age of the ONC is about 1 Myr and the average radius of its 0.4-1.2 M$_\odot$ stars is 2.1 R$_\odot$. NGC 2264 stars of the same mass (spectral class) range are significantly smaller, with a mean radius of 1.7 R$_\odot$, implying an age of about 2 Myr. ZAMS cluster stars are, of course, even smaller and individual radii can be employed in calculating j for them since their effective temperatures and luminosities are much more secure than for the PMS stars. Note that radius evolution is very non-linear over the course of the first 30 Myr; most of the contraction occurs within the initial few Myr.

It is evident, to a first approximation, from Fig. \ref{jhists} that the evolution of j occurs in a bimodal fashion. The high-j stars (i.e. the rapid rotators) evolve with essentially no loss of angular momentum from ONC age to the ZAMS clusters. In this interpretation, no physical mechanism beyond conservation of angular momentum is required to explain the evolution of about half the sample. Empirically, there is no need for angular momentum loss (by disk locking, stellar winds or any other process) for the 50-60\% of the sample that is already among the rapid rotators at ONC age. Conversely, there is no need to invoke any special mechanism to explain the fast rotation of these stars. They were already spinning rapidly at 1 Myr and they have spun even faster as they contracted to the ZAMS.

On the other hand, the initially slowly rotating stars in the ONC follow a different evolution. It is clearly seen that they continue to lose angular momentum as they evolve to NGC 2264 age and beyond. The ZAMS {\it j}-distribution is about a factor of 3-5 broader than the ONC {\it j}-distribution entirely because of the low-{\it j} stars. To summarize, we can understand to a first approximation the evolution from PMS to ZAMS in terms of only two processes: angular momentum conservation, which applies for the initially rapidly rotating half of the sample, and a braking mechanism that applies for the initially slowly rotating half. The distribution broadens on one side only -- the slowly rotating side. This is an important clue to the braking mechanism, which we discuss in Section 6.

\bigskip

\noindent
\subsection{\textbf{Very low-mass stars ($<0.4\,M_{\odot}$)}}
\label{vlm}

The two largest homogeneous samples of rotation periods, in the ONC and
NGC 2264, include about 200 periods per cluster for very low 
mass stars, an object class here defined as stars with spectral class later than M2, corresponding to
$M<0.4\,M_{\odot}$ or so, depending on the stellar models chosen. Together with smaller VLM period samples for slightly older 
objects, most notably in the $\sigma$\,Ori cluster ({\it Scholz and Eisl{\"o}ffel}, 
2004a), the $\epsilon$\,Ori cluster ({\it Scholz and Eisl{\"o}ffel}, 2005), and the much older
Pleiades ({\it Scholz and Eisl{\"o}ffel}, 2004b), they allow us to make meaningful statistical
comparisons between periods for VLM stars and their higher mass siblings.

The initial period distribution of VLM stars is well-defined by the large
samples in the ONC and NGC 2264. Periods usually range from a few hours up to 10\,d in 
both clusters, a dynamic range similar to solar-like stars. As evident in the bottom panel of Fig. \ref{Lamm}, however, slow rotators 
are clearly much rarer in the VLM regime. For example, in NGC 2264 only 4\% of 
the VLM stars have $P>10$\,d. This change is reflected in the median 
period, 
which is 3.33\,d for VLM stars in the ONC and 1.88\,d in NGC 2264, about a factor of 2 
lower than for more massive stars (see Fig. \ref{Lamm}). The period distribution 
in the slightly older clusters $\sigma$\,Ori and $\epsilon$\,Ori is, 
as shown by {\it Scholz} (2004) and {\it Scholz and Eisl{\"o}ffel} (2005), roughly 
comparable to the NGC 2264 sample, both in median period and limiting values. These similar distributions are not surprising in consideration of the similar ages estimated for the clusters.

\begin{figure}[t]
\resizebox{\hsize}{!}
{\includegraphics[angle=-90]{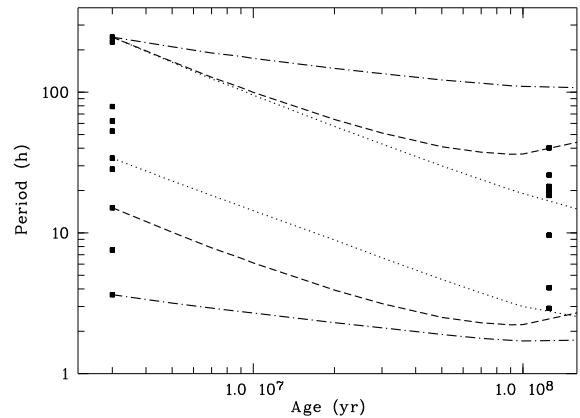}}
\caption{Rotation period of VLM stars in the $\sigma$\,Ori cluster and 
the Pleiades as a function of age (taken from {\it Scholz and Eisl{\"o}ffel}, 
2004b). The lines show model calculations, which use the periods in 
$\sigma$\,Ori as starting values. Evolutionary tracks are shown for three assumptions: angular momentum conservation (dotted lines), Skumanich-type wind losses (dashed-dotted lines) and exponential wind braking (dashed lines). }
\label{rotvlmevo}
\end{figure}

First attempts to model the angular momentum evolution in the VLM regime 
have been made by {\it Terndrup et al.} (1999) and {\it Sills et al.} (2000), both 
based on rotational velocity data of clusters with ages between 30 and 
700\,Myr. Both papers arrive at the conclusion that the rotational braking 
by stellar winds changes in the VLM regime: 
the so-called 'saturation limit' ($\omega_\mathrm{crit}$) probably drops 
quickly at very low masses, with the result that basically all VLM stars 
rotate with $\omega > \omega_\mathrm{crit}$, and are considered to be in 
the 'saturated' regime of the rotation-activity relation.
As a consequence, the rotational braking by stellar winds follows an 
exponential law ($\omega \propto \exp{(t)}$) rather than the established 
Skumanich law ($\omega \propto \sqrt{t}$) for stars of solar-like mass.

With the available VLM period data, it is now possible to have a more 
detailed look at their angular momentum evolution. Based on the VLM periods 
in $\sigma$\,Ori (age $\sim$ 3\,Myr) and the Pleiades (age $\sim$120\,Myr) Scholz and 
Eisl{\"o}ffel (2004b) investigated the rotational evolution on timescales 
of $\sim 100$\,Myr by using the $\sigma$\,Ori periods as starting points
and evolving these periods forward in time, taking into account basic 
angular momentum regulation mechanisms. That way, evolutionary tracks in 
the period-age diagram were produced which can then be compared with period 
distributions in older clusters.

Fig. \ref{rotvlmevo} shows the results from {\it Scholz and
Eisl{\"o}ffel} (2004b). Plotted are the VLM periods for $\sigma$\,Ori and the
Pleiades plus evolutionary tracks. The dotted lines show the evolution
assuming angular momentum conservation, i.e. the tracks are completely
determined by the contraction process. It is clear that this is not a
reasonable description of the available period data, because the models
predict an upper period limit $<20$\,h in the Pleiades, whereas the observed
periods range up to $\sim 40$\,h. Furthermore, it is evident from Fig.  \ref{rotvlmevo} that the most rapid rotators in  $\sigma$\,Ori would rotate at periods well below 1 hr at the age of the Pleiades (i.e. faster than the break-up velocity). We would like to point out that this upper 
period limit in the Pleiades is nicely confirmed by $v\sin{i}$ data from 
{\it Terndrup et al.} (2000). Thus, some kind of rotational braking must play a role for
VLM objects on these timescales. The second model (dash-dotted lines) includes
a Skumanich-type wind-braking law as observed for solar-type stars. In this
case, the model predicts an upper period limit $>100$\,h in the Pleiades,
significantly higher than the observed value. Thus, the Skumanich braking is
too strong and cannot be applied in the VLM regime. In a third model (dashed lines),
the model uses an exponential wind braking approach. Except for the two fastest
rotators in $\sigma$\,Ori, model and period agree fairly well. Thus, in
agreement with {\it Terndrup et al.} (2000), it was found that the period evolution
for VLM objects on timescales of $\sim100$\,Myrs is mainly determined by
contraction and exponential angular momentum loss by stellar winds.

How can we explain the failure of the best-fitting model for the fastest
rotators in $\sigma$\,Ori? First of all, the existence of these fast rotators
in $\sigma$\,Ori has been confirmed independently by {\it Zapatero-Osorio et
al.} (2003), who found a brown dwarf with a period of only 3.1\,h in this
cluster (see Section. \ref{bd}). It has to be mentioned that these objects rotate
fast in terms of their breakup period, which is around 3-5\,h for VLM stars at
the age of $\sigma$\,Ori. Their fast rotation might change the physics of
these objects, e.g., it can be expected that they are strongly oblate. It is, perhaps, 
not surprising that the simple models by {\it Scholz and Eisl{\"o}ffel},
(2004b) cannot provide a correct description of the ultrafast rotators. 
Clearly more sophisticated modeling has to be done for these  objects.

The fact that the rotation of VLM objects follows a weak exponential braking
law rather than the Skumanich law has been interpreted in terms of the
magnetic field structure of these objects ({\it Barnes}, 2003; Scholz and
Eisl{\"o}ffel 2004b). VLM objects are fully-convective throughout their
lifetime, and will never develop a radiative core. As a consequence, they are
probably not able to host a solar-type, large-scale dynamo, which is believed
to operate in the transition layer between convective and radiative zone.
Alternative mechanisms to explain the magnetic activity in the VLM regime are the
so-called turbulent dynamo ({\it Durney et al.}, 1993) and the $\alpha^2$ ({\it Chabrier and Kueker}, 2006). Both types of dynamos predict reduced Alfven radii in comparison with the solar-type $\alpha \omega$ dynamo and thus only weak braking by stellar winds. Therefore, the results from the rotational evolution
analysis are consistent with the possibility of a change of dynamo mechanism in the VLM regime.

It should be noted that the cited models are not able to constrain the
influence of more rapid angular loss mechanisms (e.g., from a disk interaction) on the rotational evolution, because they operate on
timescales of 100\,Myr. Disk-interaction timescales, for example, are probably only a
few Myr (see Section 6). To assess the validity of efficient loss mechanisms such as the disk-interaction hypothesis in the VLM
regime, it is clear that younger object samples have to be considered. This has been done by
{\it Lamm et al.} (2005), in their comparison of the period samples in the ONC and
NGC 2264. They found that the period evolution for VLM stars from $\sim1$\,Myr
(ONC) to $\sim2$\,Myr (NGC 2264) can be described with a scenario of
'imperfect' disk-locking, in the sense that the rotation of the stars is not
actually ``locked' with constant period. Instead, it spins up, but not as
fast as it would with the assumption of angular momentum conservation. Thus,
the disk brakes the rotation somewhat, but the interaction between star and
disk is less efficient than for solar-mass stars. See Section 6 for more discussion of this.

We conclude that the mechanisms of angular momentum regulation in the VLM
regime are similar to solar-mass stars, but
the efficiency of these mechanisms is a function of mass. They appear to
be less efficient in the VLM regime, resulting in more rapidly rotating objects on the ZAMS. We also note that, since the radii of the VLM stars are smaller, at all ages, than their solar-like siblings, for a given amount of specific angular momentum they spin faster. As {\it Herbst et al.} (2001) have shown, j does not vary much with mass in the ONC, so the faster rotation of the VLM stars at 1 Myr may be entirely a result of their having contracted to smaller sizes than the higher mass stars.

\subsection{\textbf{Brown Dwarfs}}
\label{bd}

In this section we discuss the available rotational data for brown dwarfs
(BDs), i.e. substellar objects intermediate in mass between stars and
planets. BDs are defined as objects with masses below the hydrogen burning
mass limit ($\sim0.08\,M_{\odot}$, {\it Chabrier and Baraffe}, 1997). Since it is
only rarely possible to determine the object mass directly, the effective
temperature and luminosity in combination with stellar evolutionary tracks is usually used 
to identify BDs. Therefore, in the following we refer to `brown dwarfs' as
objects whose spectral types and luminosities classify them as substellar, although some of
them might have masses slightly higher than the substellar limit, because of
uncertainties in spectral typing and atmosphere modeling. 

Following the discovery of the first BDs ({\it Nakajima et al.}, 1995, {\it Rebolo et al.}, 1995), hundreds of them have been identified in star forming regions
(e.g., {\it Comer\'on et al.}, 2000, {\it L\'opez Mart{\'\i} et al.}, 2004), open clusters 
(e.g., {\it Zapatero-Osorio et
al.}, 1997, {\it Barrado y Navascu\'es et al.}, 2001), and in the field ({\it Kirkpatrick et
al.}, 1999, {\it Phan-Bao et al.}, 2001). A large number of follow-up studies led to
rapid progress in our understanding of the physical properties of
these objects. For example, about 30 rotation periods have been measured for
BDs at different ages, complemented by rotational velocity ( {\it v} sin {\it i}) data. Here, we review the available periods and their implications for our
understanding of rotation and the angular momentum evolution of brown dwarfs.

The first rotation period for a likely BD 
was published for an object in the open cluster $\alpha$\,Per, which
has an age of $\sim$50\,Myr ({\it Mart\'{\i}n and Zapatero-Osorio}, 1997). In the
following years, periods have been measured for three BDs in the $\sim$1\,Myr old
Cha I star forming region ({\it Joergens et al.}, 2003), three objects in the $\sim$3 Myr
old $\sigma$\,Ori cluster ({\it Bailer-Jones and Mundt}, 2001, {\it Zapatero-Osorio et
al.}, 2003), one object in the $\sim$120\,Myr old Pleiades cluster ({\it Terndrup et al.}, 1999), and
three so-called 'ultracool dwarfs' in the field ({\it Bailer-Jones and Mundt}, 2001,
{\it Clarke et al.}, 2002), with late-M or early-L spectral types. 
We would like to caution that it is not, in all cases, unambiguously clear that the
photometrically derived period corresponds to the rotation period. This is particularly
important for the ultracool field dwarfs which, in many cases, have relatively sparse
sampling and, perhaps, less stable surface features, factors that hamper a reliable period detection. In this review, we include discussion
only of periods that authors themselves consider to be likely 
rotation periods.

\begin{figure}[t]
\resizebox{\hsize}{!}
{\includegraphics[angle=-90]{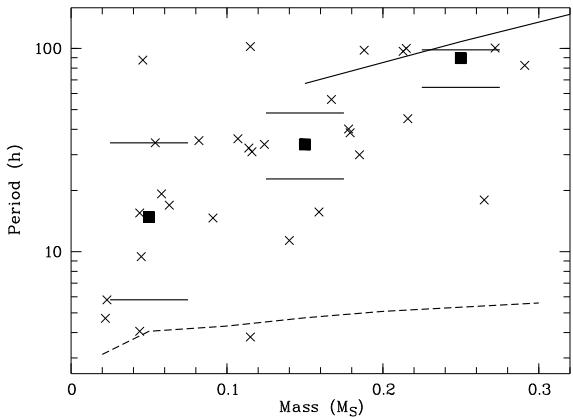}}
\caption{Period vs. mass for the $\epsilon$\,Ori cluster (figure from Scholz
  and Eisl{\"o}ffel 2005). Filled squares mark the period median, horizontal
  lines the quartiles. The solid line is the period median for the ONC, the
  dashed line the breakup limit for $\epsilon$\,Ori.}
\label{permasscrit}
\end{figure}

It is interesting to note that until 2003 all known periods for objects with
ages $>1$\,Myr were shorter than one day, providing the first evidence that evolved
BDs are, in general, very rapid rotators. This is supported by spectroscopic
rotational velocity data, indicating that the majority of evolved BDs have 
$v\sin i > 10\,\mathrm{km s^{-1}}$ ({\it Mohanty and Basri}, 2003; {\it Bailer-Jones}, 2004). 
By contrast, the rotation periods for the three youngest objects in the sample, i.e. those in Cha I, 
are in the range of 2 to 4 days.

In the last two years, deep wide-field monitoring campaigns have more than
doubled the number of known periods for BDs. In the young open clusters
$\sigma$\,Ori and $\epsilon$\,Ori, {\it Scholz and Eisl{\"o}ffel} (2004a, 2005)
measured photometric periods for 18 probably substellar objects, 9 for each
cluster. Both clusters belong to the young population of the Ori OB1b
association, which has an age of about 3\,Myr, although the $\sigma$\,Ori
objects are on average probably slightly younger than those in $\epsilon$\,Ori
({\it Sherry}, 2003). In $\sigma$\,Ori, the periods cover a range from 5.8 to 74\,h
with a median of 14.7\,h, whereas in $\epsilon$\,Ori the total range is 4.1 to
88\,h with a median of 15.5\,h. Thus, the BD periods in both clusters are
comparable. Additionally, periods for two likely substellar members of the
Pleiades ({\it Scholz and Eisl{\"o}ffel}, 2004b) and IC4665 (age $\sim$ 40\,Myr, {\it Scholz}, 2004)
have been published. In total, the rotation sample for BDs (or objects very
close to the substellar limit, see above) comprises 31 periods. In the
following, we will discuss the period-mass relationship and rotational evolution
in the substellar regime based on this dataset.

To separate age and mass effects, the period-mass relationship has to be
studied for each age separately. In Fig. \ref{permasscrit} (taken from {\it Scholz
and Eisl{\"o}ffel}, 2005) period vs. mass is plotted for the $\epsilon$\,Ori
objects (crosses), where one of the largest BD period samples is
available. The figure additionally shows the median period for certain mass
bins (filled squares), together with the period-mass relationship in the
ONC (solid line). We have already demonstrated in Sectiion. \ref{vlm} 
that the average period decreases with decreasing object mass in the very low mass
star regime. As can be seen in Fig. \ref{permasscrit}, this trend continues
well-down into the substellar regime. The median in the BD regime is clearly
lower than for VLM stars. The same result is obtained for the period sample in
the $\sigma$\,Ori cluster. It is particularly interesting that the BD period
range in young open clusters extends down to the breakup period, which is the
physical limit of the rotational velocity. In Fig. \ref{permasscrit} the
breakup limit is over-plotted as dashed line (calculated using the evolutionary
tracks of {\it Baraffe et al.}, 1998). For masses below $0.1\,M_{\odot}$ a sub-sample
of objects rotates very close to breakup. Since extremely fast rotation
affects the evolution of the objects, as we know from massive stars, this has
to be taken into account in future evolutionary models for rapidly rotation BDs.

\begin{figure}[t]
\resizebox{\hsize}{!}
{\includegraphics[angle=-90]{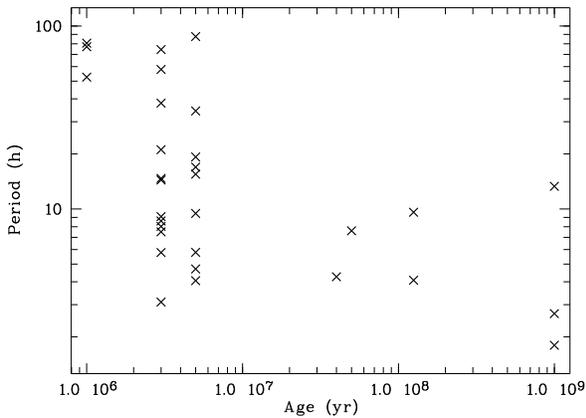}}
\caption{Rotation period of brown dwarfs (and objects very close to the
  substellar limit) as a function of age. For simplicity, periods for evolved
  field dwarfs were all plotted at an age of 1\,Gyr.}
\label{rotbdevo}
\end{figure}

All available BD periods are plotted in Fig. \ref{rotbdevo} as a function of
age. For the matter of simplicity, we assigned an arbitrary age of 1\,Gyr to
all evolved field objects. Whereas very young BDs have periods ranging from a few hours
up to four days, all periods for older objects are shorter than 15\,h. Please
note that the upper period limit in the youngest clusters may be set, in part, by the upper period detection limit. In the $\sigma$\,Ori cluster, however,
the detection limit is $\sim 10$\,d ({\it Scholz and Eisl{\"o}ffel}, 2004a), leading
us to the tentative conclusion that 100\,h might be a realistic value for the
upper period limit at very young ages. Two clear trends can be seen from
Fig. \ref{rotbdevo}, related to the upper and the lower period limit, which we
will discuss separately in the following paragraphs.

First, the upper period limit is apparently more or less constant for ages $<5$\,Myr and is
decreased by a factor of about 6-10 in the more evolved clusters with ages
$>40$\,Myr. 
The second aspect of Fig. \ref{rotbdevo} that we wish to discuss is the lower period limit, which appears to be in the range of a few hours at
all evolutionary stages. The only exception is the (very sparse) period sample
for Cha I at 1\,Myr, but this might be related to small number statistics or a
time series sampling unable to detect short periods. Within the statistical
uncertainties, the lower period limit is constant with age. Thus, for a
fraction of ultrafast rotating BDs the period changes by less than a factor of
two on timescales of $\sim$1\,Gyr. This is surprising, because on the same
timescales BDs contract and we should therefore expect a rotational
acceleration at least by a factor of about ten. One possible explanation is
that we have not yet found the fastest rotators among the evolved
BDs. Assuming angular momentum conservation, we should expect objects with
periods down to $\sim$0.5 h at ages $>200$\,Myr, i.e. when the contraction
process is finished. Whether these objects exist or not, has to be probed by
future observations.

If the lower period limit for evolved BDs, however, is really in the range of
a few hours, as indicated by the available period data, the fastest rotators
among the BDs have to experience strong angular momentum loss on timescales of
1\,Gyr. It is unclear what mechanism could be responsible for this rotational
braking. To summarize this section, while the rotation data on BDs are still very scanty compared to the low mass stars, they appear to be a natural extension of the phenomena observed for stars. There is nothing yet to suggest that their rotation properties and evolution are discontinuous in any way from stars.

\bigskip

\section{\textbf{Overview of Variability, Spots, Accretion and Magnetic Star-Disk Interactions in YSOs and BDs}} 
\bigskip

Most of our knowledge of the rotational properties of the
objects discussed here is based on variability studies.
Therefore we regard it as important to give an overview on
this subject, with particular attention
to magnetically induced cool stellar spots 
and magnetically channeled variable mass accretion, the principle variability mechanisms in weak (WTTS) and classical (CTTS) T Tauri stars, respectively.

A detailed study of the various sources of TTS variability has 
been carried out by {\it Herbst et al.} (1994), based on a large electronic 
UBVRI catalog with about 10,000 entries for several hundred stars.
A further variability study, with particular
attention to periodic variations, has been carried out by
{\it Bouvier et al.} (1995).
On the basis of these two and related studies one can distinguish at least
5 types of common PMS variability, at least the first two of which are also seen in BDs. These are:

1. Periodic variability caused by rotational modulation of the stellar flux
by an asymmetric distribution of cool spots or spot groups on
the stellar surface. This type of variability is more frequently seen in  WTTSs  but
can also be observed in the CTTSs. An example is shown in Fig. \ref{star019}. The typical amplitudes
for these variations range from about 0.03-0.3 mag in the V band,
with the most extreme values reaching 0.8 mag in V and
0.5 mag in I. Spot sizes and temperatures have been derived from the observed 
amplitudes and the derived spot coverage factors range from a few percent up to 30$\%$ (for $\Delta$V=0.5 mag, see {\it Bouvier
et al.}, 1995). 

2. Irregular variations, probably caused by highly variable, 
magnetically channeled accretion from the circumstellar disk onto the star. 
The accretion rate onto the star is not only variable in time 
but the accretion zones are certainly not uniformly distributed over the stellar surface.
The complex interaction between the stellar 
magnetosphere and the inner disk is evidently highly dynamic and time dependent.
The typical amplitudes of the resulting (largely) irregular variations are a factor of 2-5 larger in
V than those of the periodic variations observed in many WTTs. 
Variations by 1.5 mag in V within a few days are not unusual and some stars can vary that much within hours. This type of variability is designated as Type II by {\it Herbst et al.} (1994).

3. Periodic variations due to hot spots. This type of variation (also known as Type IIp)
is only seen in CTTS and the hot spots are presumably at the base of the magnetic channels. The periodicity typically persists for only a couple of rotation cycles.  Since the
magnetic field configuration is highly unstable, the size and location
of these spots is correspondingly changing within a few rotation periods.
This is quite in contrast to the cool spots which may last for hundreds
to thousands of rotations. The amplitudes of the rotational modulation
by hot spots is typically a factor 2-3 larger in V than those seen in WTTS
due to cool spots, but more extreme cases have been observed in some CTTSs.

4. Flare-like variations, mainly seen in the U and B-band in the 
WTTSs. This type of variability is probably also  present in CTTS,
but difficult to distinguish from the strong irregular variability. 

5. UX Ori-type variability (also referred to as UXors) is mostly seen in early type TTS
(earlier than K0) and in Herbig Ae/Be stars. It is designated as Type III variability by {\it Herbst et al.} (1994). The variation amplitudes can
be very large (up to 2.8 mag in V), but the time scales
are about a factor of 2-10 longer compared to the irregular variations
of CTTS. Also, the stars often get bluer when fainter and the H$\alpha$ flux does not correlate with the continuum flux as it does in the CTTS. {\it Herbst and Shevchenko} (1999) discuss this type of variability in detail. Its cause is still uncertain although many authors believe it derives from variations in circumstellar extinction. 

The photometric behavior of  numerous periodically variable 
WTTS on timescales of several years has been studied in the ONC by {\it Choi and Herbst} (1996) and {\it Steinhauer et al.} (1996) and in IC\,348 by {\it Cohen et al.} (2004) and  {\it Nordhagen et al.} (in preparation). In the latter
case the study now extends over 7 years.
These investigators found clear variations
in the amplitude and light-curve shape of these periodic variables
on time scales of less than one year, probably due to changes in the
spot sizes and spot distributions (see Fig. \ref{star019}). In no case, however, did they find definitive
evidence for a change in period by more than the measurement limit
of about 1$\%$, indicating that differential rotation in WTTSs is much
less than in the Sun or that spots are confined to certain latitude zones. This conclusion is also supported by the very similar
rotation period values found in the ONC by {\it Herbst et al.} (2002) in
comparison to previous studies by {\it Stassun et al.} (1999) and {\it Herbst et al.},
(2000a). In NGC\,1333 there may be an ``exception that proves the rule''.
A T Tauri star in that region has been found with a period of 5.6 d over three seasons and then 4.6 days over the next two seasons, a change exceeding 20$\%$ ({\it Herbst et al.}, 2005). This proves that such stars can be found, but they are clearly exceedingly rare.

\begin{figure}[t]
\resizebox{\hsize}{!}
{\includegraphics[angle=0]{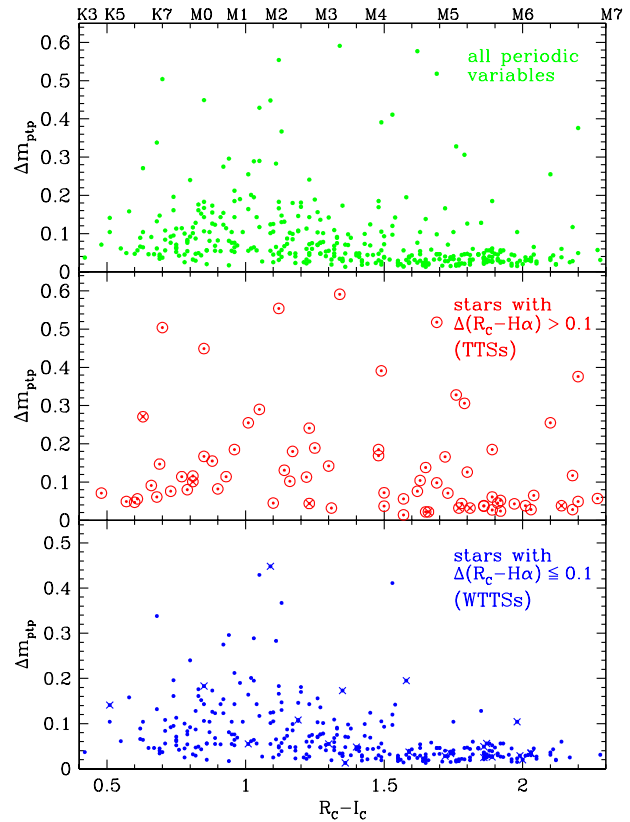}}
\caption{The \ptp variation of the 405 periodic 
                variables in NGC\,2264 
                found by {\it Lamm et al.} (2004) as a function of their R-I color.
                In the top panel all 405 periodic variables are 
                shown. The diagram in the middle panel contains only 89
                stars with strong H$\alpha$ emission.
                In the bottom panel only the 316 stars with weak H$\alpha$ 
                emission (WTTSs) are shown. Note the strong 
                decrease in amplitude, by about
                a factor 3 for the WTTS in the bottom panel with R-I$\ge$1.5 (from
                {\it Mundt et al.}, in preparation).}
\label{ptp}
\end{figure}

It is apparent in the data that the amplitudes due to rotational modulations
by  cool spots strongly decrease with age. While  WTTS commonly have spot  amplitudes
of 0.1-0.3 mag in V, this
drops to about 0.02 mag for stars in the Hyades (600\,Myr) and at least another
factor of 10 in the case of our Sun ({\it Strassmeier}, 1992). The same is true
for VLM stars. While the amplitudes for the VLM stars in NGC\,2264 (see Fig.~\ref{ptp})
are similar to those observed in the Pleiades ({\it Scholz and Eisl\"offel}, 2004b --
which could be due to biasing of detections towards large amplitudes)
they drop by about a factor 3 for late field M stars ({\it Rockenfeller et al.}, in preparation).
It is obvious that the decreasing amplitudes with increasing age makes it more
difficult to study photometrically the rotational properties of ZAMS and older clusters.

The dependence of amplitudes of the periodic light variations in PMS stars on 
mass (or effective temperature) has been investigated by {\it Mundt et al.} (in preparation)
for the case of NGC\,2264 over a broad range of spectral types
from $\sim$K3 to $\sim$M6.5. This investigation is based on the
the data of {\it Lamm et al.} (2004, 2005). The main results are 
illustrated in Fig.~\ref{ptp},
which  shows the \ptp variation as a function of the 
$\ri$ color for three different sub-samples of the periodic variables.
The \ptp amplitudes were derived by fitting a sine wave to
the phased light curves. In the top panel of Fig.~\ref{ptp}, all 405 periodic 
variables are shown.  
In the middle panel  only the \ptp variations of the 89 stars with large 
\haindices ($\hain\,>\,0.1\mag$, i.~e. strong $\hamag$ emitters) are shown. 
It is evident that these stars show a large scatter in their 
\ptp variations, i.e. 
we probably deal with a mixture of stars in which either cool or hot spots
are responsible for the periodic variability.
Most important for our discussion is the bottom panel of Fig.~\ref{ptp}, 
which shows the \ptp variations for the remaining 316 stars which are 
mostly WTTSs due to their   weak $\hamag$ emission 
(i.~e. $\hain\,\leq\,0.1\mag$). From this panel 
it is clearly evident that the \ptp variations of the cooler WTTSs 
with $\ri\,\ga\,1.5\mag$ show on average a factor $\sim$3 smaller
\ptp variations than stars with $\ri\,\la\,1.3\mag$.
This impressive decrease in the \ptp variations for stars with 
$\ri\,\ga\,1.5\mag$ is practically independent of the period of the 
investigated stars. Only for the slowest rotators might there be a 
tendency for somewhat higher \ptp variations.
We note that  the objects with $\ri\,\ga\,2.0\mag$ are all VLM stars
($\le$0.1-0.15$\msun$) with some of
them probably falling below the substellar limit. A similar decrease in
the \ptp variations of the periodic variables with decreasing mass was found
by {\it Scholz and Eisl\"offel} (2004) in the Pleiades.

This tremendous decrease in the \ptp variations of the coolest periodically variable WTTSs 
either implies that the  spot coverage of the stellar surface, the asymmetry
of the spot distribution, and/or the contrast between the spot and the photospheric
environment has decreased. It has been argued by {\it Mundt et al.} (in preparation) that 
the spot coverage (spot size) has probably decreased as a result of the much poorer coupling between
the magnetic fields and the atmospheric plasma caused by the low ionization fraction in the atmosphere of these very cool objects.
The change in \ptp variations of WTTSs due to cool spots was also investigated 
by {\it Bouvier et al.} (1995) but only for spectral types between about G0 and
K7 and only for a sample of 23 stars. In this spectral range the \ptp variations 
apparently {\it increase} with decreasing temperature. Such a behavior is not inconsistent
with the data displayed in  Fig.~\ref{ptp}, since there the median value for the
\ptp variations, in fact, increases until  $\ri\,\sim\,1\mag$ ($\sim$M1).

It is quite obvious from Fig.~\ref{ptp} and  the work summarized above 
that periodic light modulations due to cool spots are observable over a very
large mass range; i.e. from about 1.5\,M$_{\odot}$ well down into the substellar mass regime.
We know that the same is true for variable mass accretion and 
accretion related activity phenomena. Such CTTS-like phenomena are observable in young VLM stars
and young BDs down to masses near the deuterium burning limit (see e.g.,
{\it Mohanty et al.}, 2005a for a review). Examples of such accretion-related activity are numerous and include
strong line-emission (e.g., {\it Mohanty et al.}, 2005b), variable line-emission (e.g., {\it Barrado
y Navascu{\'e}s et al.}, 2003), irregular variations in the continuum flux (e.g., 
{\it Zapatero-Osorio et al.}, 2003, {\it Scholz and Eisl\"offel}, 2004a, 2005), and bipolar emission 
line jets ({\it Whelan et al.}, 2005). It remains for us to discuss the confrontation between theory and observation which involves the connection between rotation, magnetic fields and accretion disks in all of the mass ranges discussed here from stellar to substellar.

\section{Interpretations: Comparison of Empirical Results with Theory}

The leading theory to account for the slower than critical rotation rates observed for T Tauri stars is commonly known as ``disk locking" and was first proposed by {\it Camenzind} (1990) and {\it K\"onigl} (1991). It has been worked out in more detail by {\it Shu et al.} (1994) and others, most recently by {\it Long et al.} (2005). Its principal success in the rotation arena  is in predicting an equilibrium rotation period in the 2 to 8 day range, characteristic of the observations, when other parameters, such as surface magnetic field strength and accretion rate, are set to nominally representative values.  Observational credence is given to this picture by the fact that a statistically significant anti-correlation does exist between angular velocity and various disk indicators such as near-infrared excess and H$\alpha$ equivalent width ({\it Edwards et al.}, 1993; {\it Herbst et al.}, 2002; {\it Lamm et al.}, 2004, 2005; {\it Dahm and Simon}, 2005; {\it Rebull et al.}, 2005). Fig. 8 shows the anti-correlation for one cluster, the ONC.

\begin{figure}[t]
\resizebox{\hsize}{!}
{\includegraphics{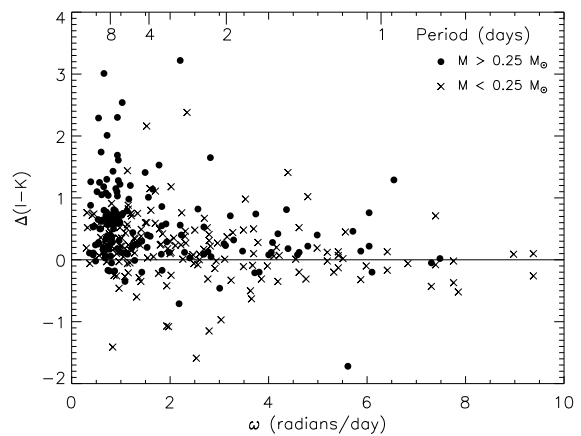}}
\caption{Infrared excess emission depends on rotation for PMS stars in the ONC (from {\it Herbst et al.}, 2002). The indicated mass range is based on a spectral class range and would translate into masses greater than 0.4 solar masses by some models. Since infrared excess is an indication of an accretion disk, this figure clearly demonstrates the rotation-disk connection.}
\label{irexcess}
\end{figure}

Note that these anti-correlations, including the one in Fig. 8, generally indicate that rapid rotators are very unlikely to have disks, while slow rotators may or may not have them. This is actually what one would expect from a disk locking (or, more generally, disk regulation) scenario. In particular, a star does not respond instantaneously to the loss of angular velocity regulation. It takes time for a star to contract and spin up. Some stars that are no longer regulated, i.e. have lost their accretion disks recently, would be expected not  to have had time yet to spin up. They would appear as slow rotators that lacked accretion disks. It would be harder, in this scenario, to explain rapid rotators with disks and, indeed, such stars are rare in the sample. It would be interesting to inquire more deeply if there are, in fact, any such cases that cannot be ascribed to errors of observation.

In spite of its successes, the disk locking theory has been controversial over the years. Many authors have pointed to various shortcomings of either the theory or its confrontation with observation (e.g., {\it Stassun et al.}, 1999, 2001; {\it Rebull}, 2001; {\it Bouvier et al.}, 2004; {\it Uzdensky}, 2004; {\it Littlefair et al.}, 2005; {\it Matt and Pudritz}, 2004, 2005a,b). The interested reader is referred to the meeting report by {\it Stassun and Terndrup} (2003) and references therein, as well as to the recent review by {\it Mathieu} (2004) for further discussion. Here we note that the difficulties appear to us to be rooted in the natural complexity of the phenomena and our limited ability to either model them or to obtain a sufficient amount of accurate data to empirically constrain them. As with all MHD processes, theoretical progress can only be made with simplifying assumptions such as axisymmetric magnetic fields and steady accretion, which we know in the case of T Tauri stars are not realistic. For example, if the geometry of the problem were truly axisymmetric for a typical star then we would not observe cyclic photometric variations with the rotation period. If accretion were truly steady, we would not see the large amplitude irregular photometric variations characteristic of CTTS.  

Time-dependent magnetic accretion models
may result in a decreased braking of the stellar rotation rates
in comparison to the simple disk locking scenario 
({\it Agapitou and Papaloizou}, 2000). It has been argued by {\it Matt and Pudritz}, 
(2005a,b) that the magnetic braking of PMS stars may not be due to a magnetic star-disk
interaction, but may result from a magnetically driven wind emanating directly
from the star. If the high rotation rates of Class\,I protostars compared to
Class\,II protostars (CTTSs) observed by {\it Covey et al.} (2005) are confirmed 
by further studies they may imply angular momentum loss rates higher 
than predicted by the disk locking scenario. Note that in the disk locking
scenario the Class\,II sources should rotate as fast as the Class\,I sources.
If they in fact rotate a factor of 2 slower, it may demand a breaking
mechanism more efficient than disk locking to account for the observations.
A strong magnetically driven wind, as proposed by {\it Matt and Pudritz} (2005a,b), 
is certainly an idea which deserves further study. An observational
argument in favor of a wind emanating directly from the stellar surface is
the broad and deep P Cygni profiles observed in some CTTSs (see e.g., {\it Mundt}, 1984).
The deep blue-shifted absorptions of these line profiles would be much harder to understand if the wind acceleration region is far from the stellar surface, as would be the case for a disk wind.

From the theoretical side, part of the difficulty in testing the disk locking theory is that it is hard to pin down specific, testable predictions that are based on realistic models, especially when parameters such as the magnetic field strength are involved, that are hard to observe. Some attempts to test the theory in detail have met with mixed success ({\it Johns-Krull and Gafford}, 2002). Observationally, we are further faced with the problem of a very broad distribution of rotation rates at any given mass and age as well as the difficulties of even establishing mass and age for PMS stars. Magnetic phenomena are notoriously complex and this apparition of their importance is no exception. Considerations such as these have led {\it Lamm et al.} (2004, 2005) for example to employ the terminology ``imperfect" disk-locking to account for the data. The notion that a star's rotation is fully controlled by its interaction with a disk during all of its PMS phase in a way that can accurately be described by a current disk-locking theory is probably oversimplified. Nonetheless, the concept of disk regulation of, or at least effect on, rotation seems undeniable in the light of the observed correlations, such as seen in Fig. 8. 

Indeed, there is no way to understand the evolution of rotation of stars from PMS to ZAMS without invoking significant braking for a significant amount of time (c.f. Figure 3). All attempts to model this have employed such ``disk locking" (e.g., {\it Armitage and Clarke} 1996; {\it Bouvier et al.}, 1997;  {\it Krishnamurthi et al.}, 1997; {\it Sills et al.}, 2000; {\it Barnes et al.}, 2001; {\it Tinker et al.}, 2002; {\it Herbst et al.}, 2002; {\it Barnes}, 2003; {\it Rebull et al.}, 2004; {\it Herbst and Mundt}, 2005). Usually, the approach is to include the ``disk locking time" as a parameter in the models. In its simplest form this is the time that the period remains constant, after which it is allowed to change in response to the contraction of the star. 

Obviously a real star would not be expected to maintain a constant rotation period for times of order 1 Myr, even if the disk locking theory were strictly true, since other parameters of the problem, such as magnetic field strength and accretion rate are likely to vary on this time scale (or shorter -- perhaps much shorter). Hence, the parameterized models are only approximations of reality. What is interesting and significant is that it is simply not possible to model the rotational evolution of solar-like (0.4-1.2 M$_\odot$) stars from PMS to ZAMS without significant braking for about half of them. And, the observations show that the half that must be braked is the half already slowly rotating (Figure 3), precisely the same stars that are most likely to show evidence of circumstellar disks (Figure 8). We note that the maximum braking times found by various authors are of the same order, normally around 5-10 Myr, comparable to the maximum lifetime of accretion disks derived from near-infrared studies (e.g., {\it Haisch et al.}, 2001). Of course, it is possible that the influence of a disk on rotation may wane before its detectability in the infrared, so if disk-locking times are somewhat shorter than disk-detectability times this may not be surprising.  Nonetheless, rotation studies and infrared excess emission studies appear to concur in indicating that substantial gaseous accretion disks have disappeared by about 5 Myr for almost all stars and by about 1 Myr for half of them. This, in turn, implies that the era for gas giant planet formation has ended. Terrestrial planets may, of course, continue to form around such stars for much longer periods of time. 

Finally, we would like to make some comments on the disk locking
scenario and on magnetic star-disk coupling in general as it applies to VLMs and BDs. As discussed above, the 
disk locking scenario is quite successful, in some ways, in explaining important aspects 
of the rotational evolution of low-mass stars with masses larger than about
0.4\,M$_{\odot}$. On the other hand for VLM stars below this mass 
limit the evidence is less convincing (see e.g., {\it Lamm et al.}, 2005). Nevertheless, it appears
from the available data that those VLM objects showing evidence
for active accretion (e.g., strong H$_{\alpha}$ emission) do rotate on average
much slower than those stars without any accretion indicators.
This means that these low-mass objects do indeed lose angular momentum,
but probably at a lower rate than in the case of precise disk locking. Therefore the
term ``moderate angular momentum loss'' was proposed by ({\it Lamm et al.}, 2005).

The efficiency of disk locking is likely to
decrease in the VLM star and BD regimes. Unfortunately, it is very difficult to test these ideas observationally because of the difficulties of determining rotation periods, masses and ages for large samples of VLMs and BDs. In particular, we lack a sufficient amount of data for ZAMS VLMs that could provide a goal for the PMS evolution analogous to what is available for the solar-type stars. As argued in Section 4.2, rotational braking by stellar winds is also probably less efficient for VLM objects, either
because they have no solar-like dynamo or because of little coupling between
gas and magnetic field due to low gas temperatures in the atmosphere. As
already noted above, the period distribution at very young ages extends down
to the breakup limit (see Fig. \ref{permasscrit}). This might have an effect
on the fastest rotators, in the sense that strong centrifugal forces remove
angular momentum and thus brake the rotation, as argued by {\it Scholz and
Eisl{\"o}ffel} (2005). If and how this can explain the observed period
evolution has to be investigated with detailed future modeling.

Even in the substellar regime there is some evidence for disk braking or some other form of angular momentum loss related to disks. Accreting, very young objects are nearly exclusively slow rotators, whereas non-accretors cover a broad range of rotation rates ({\it Scholz and Eisl{\"o}ffel},
2004a, {\it Mohanty et al.}, 2005a). The magnetic braking of the  disk in the PMS phase is probably less strong
than for solar-like stars, but it still might be able to 
prevent the objects from conserving angular momentum during the first few Myrs of
their evolution. After that, there seems to be very little chance for significantly slowing the rotation of BDs and their rotational evolution is probably dominated by conservation of angular momentum as they continue to contract. Since the radii of substellar objects are expected to
decrease by a factor of $\sim$3, due to hydrostatic contraction,
on timescales of $\sim$200\,Myr ({\it Baraffe et al.}, 1998), the periods should decrease
by a factor of $\sim$9 on the same time scale, assuming angular momentum conservation. Thus, the
decline of the upper period limit between 5\,Myr and the older clusters seen on Fig. 6, is probably just reflecting conservation of angular momentum and is roughly consistent with it quantitatively. 
Finally, while at young ages,
substellar objects are clearly able to maintain magnetic activity which should lead to mild rotational braking by stellar winds on timescales of
about 100\,Myr, they seem to lose their magnetic signatures as they age and cool.  Objects with spectral type L, into which these objects evolve, have been found to be
too cool to maintain significant chromospheric and coronal activity ({\it Mohanty and
Basri}, 2003). Hence, there is probably little or no rotational braking
by stellar winds on very long timescales, explaining why the upper period
limit on Figure 6 stays at a very low levels for a very long time. Again, it will clearly be necessary to expand the data sample in order to test these ideas more rigorously.

To summarize, in the time since Protostars and Planets IV we have seen a tremendous growth, by a factor of 10 or more, in the number of PMS stars for which we know the rotation period. Coupled with similar data for young clusters we now have a good, statistical picture of the evolution of surface angular momentum for solar-like stars from 1 Myr to the ZAMS (Fig. 3). To a first approximation, at 1 Myr the stars already divide into a slower rotating half and a more rapidly rotating half. The division is exaggerated over the next few Myr as the slower rotating stars continue to suffer substantial rotational braking while the faster rotators spin up in rough agreement with angular momentum conservation as they contract towards the ZAMS. The braking is disk related (see Figure 8) and persists for around 5-6 Myr. 

At the same time we have begun to probe into the low mass and substellar mass regimes with the same photometric technique, although the amplitudes of the variations (Figure 7) and faintness of the objects make it a more difficult problem to find rotation periods. Nonetheless, the data show similar kinds of behavior for these stars (Figure 5), although braking by both disks and winds appears to become increasingly less efficient as one progresses to smaller mass objects (Figure 6). It is in this mass regime that we expect the next five years to bring particular progress, since there is so much to be done. It is also likely that the improved data on disks coming from the {\it Spitzer Infrared Telescope} and elsewhere will help sharpen the observational tests relevant to disk locking and other theories. Finally, the difficult problem of what happens during the first 1 Myr (i.e. the proto-stellar and early PMS phases) to produce such a broad rotational distribution already in the ONC will hopefully become clearer as data on the highly embedded objects continue to accumulate.



\bigskip

\noindent
\textbf{ Acknowledgments.} W.H. gratefully acknowledges the continued support 
of NASA through its Origins of Solar Systems Program and the support and hospitality of the staff of the Max-Planck-Institute for Astronomy in Heidelberg during extended visits. J.E. and A.S. were partially supported by the Deutsche Forschungsgemeinschaft grants Ei 409/11-1 and 11-2. We thank the referee, Keivan Stassun, for his helpful comments on the first draft of this chapter.

\bigskip

\centerline{\textbf{REFERENCES}} 
\bigskip
\parskip=0pt
{\small
\baselineskip=11pt


\refs Agapitou V. and Papaloizou J. C. B. (2000)  {\it Mon. Not. R. Astro. Soc., 317}, 273-288.

\refs Armitage P. J. and Clarke, C. (1996) 
{\it Mon. Not. R. Astron. Soc., 280}, 458-468.

\refs 
Attridge J. M. and Herbst, W. (1992) 
{\it Astrophys. J., 398}, L61-L64.

\refs Bacciotti F., Ray T. P., Mundt R., Eisl\"offel J., and Solf J.
      (2002)  {\it Astrophys. J., 576}, 222-231. 
 
 \refs Bailer-Jones C. A. L. (2004)
{\it Astron. Astrophys., 419}, 703-712.
 
\refs Bailer-Jones C. A. L. and Mundt R. (2001)
{\it Astron. Astrophys., 367}, 218-235.

\refs Baraffe I., Chabrier G., Allard F., and Hauschildt P. H. (1998)
{\it Astron. Astrophys., 337}, 403-412.

\refs Barnes S. A. (2003)
{\it Astrophys. J., 586}, 464-479.

\refs Barnes,S. A., Sofia S., and Pinsonneault, M. (2001)
{\it Astrophys. J., 548}, 1071-1080.

\refs Barrado y Navascu\'es D., Stauffer J. R., Briceno C., Patten B., Hambly N. C., and
Adams J. D. (2001) {\it Astrophys. J. Supp., 134}, 103-114.

\refs Barrado y Navascu{\'e}s D., B{\'e}jar V. J. S., Mundt R., Martin E. L., Rebolo R., 
Zapatero-Osorio M. R., and  Bailer-Jones C. A. L. (2003) {\it Astron. Astrophys., 404}, 171-185.

\refs Bodenheimer P. (1995) 
{\it Ann. Rev. Astron. Astrophys., 33}, 199-238.

\refs 
Bouvier J. (1994) in {\it Cool Stars; Stellar Systems; and the Sun; Eighth Cambridge Workshop. Astronomical Society of the Pacific Conference Series, 64}, (J.-P. Caillault, ed.),  pp. 151-158, Astron. Soc. of the Pacific, San Francisco.

\refs 
Bouvier J. and  Bertout C. (1989) 
 {\it Astron. Astrophys., 211}, 99-114.

\refs 
Bouvier J., Bertout C., Benz W., and Mayor M. (1986) 
 {\it Astron. Astrophys., 165},110-119.

\refs 
Bouvier J., Cabrit S., Fernandez M., Martin E. L., and Matthews J. M. (1993) 
 {\it Astron. Astrophys. Supp., 101}, 485-505.
 
\refs 
Bouvier J., Covino, E., Kovo, O., Martin, E. L., Matthews, J. M., Terranegra, L., and Beck, S. C.  (1995) 
 {\it Astron. Astrophys., 299}, 89-107.

\refs 
Bouvier J., Forestini M., and Allain S. (1997)
 {\it Astron. Astrophys., 326}, 1023-1043.
 
 \refs Bouvier J., Dougados C., and Alencar S. H. P. (2004)
 {\it Astrophys. and Space Sci., 292}, 659-664. 

\refs Camenzind M (1990) {\it Rev. Modern Astron., 3}, 234-265.

\refs 
Carpenter J. M., Hillenbrand L. A., and Skrutskie M. F. (2001) 
{\it Astron. J., 121}, 3160-3190. 

\refs Chabrier G. and  Baraffe I. (1997)
{\it Astron. Astrophys., 327}, 1039-1053. 

\refs Chabrier G. and Kueker L. (2006) {\it Astron. Astrophys., 446}, 1027-1038.

\refs Choi P. I. and Herbst W. (1996) {\it Astron. J., 111}, 283.

\refs Clarke F. J., Tinney C. G., and Covey K. R. (2002) {\it Mon. Not. R. Astron. Soc. 332}, 361-366.

\refs Comer\'on F., Neuh{\"a}user R., and Kaas A. A. (2000)
{\it Astron. Astrophys., 359}, 269-288.

\refs Cohen R. E., Herbst W., and Williams E. C. (2004) 
{\it Astron. J., 127}, 1594-1601. 

\refs 
Covey K. R., Greene T. P., Doppmann G. W., and Lada C. J. (2005) 
 {\it Astron. J., 129}, 2765-2776.

\refs 
Dahm, S. and Simon, T. (2005)
{\it Astron. J., 129}, 829-855.

\refs Durney B. R., De Young D. S., and Roxburgh I. W. (1993)
{\it Solar Physics, 145}, 207-225.

\refs Edwards S., Strom S. E., Hartigan P., Strom K. M., Hillenbrand L. A., Herbst W., Attridge J., Merrill K. M., Probst R., and Gatley I. (1993) {\it Astron. J. 106}, 372-382.

\refs Haisch K. E., Lada E. A.' and Lada C. J. (2001)
{\it Astrophys. J., 553}, L153-156.

\refs Hartmann L., Hewett R., Stahler S., and Mathieu R. D. (1986)
{\it Astrophys. J., 309}, 275-293.

\refs Herbst W. and Shevchenko V. S.. (1999)  {\it Astron. J., 118}, 1043.

\refs Herbst W. and Mundt R. (2005)  {\it Astrophys. J., 633}, 967-985 .

\refs Herbst W., Booth J. F., Chugainov P. F., Zajtseva G. V., Barksdale W., Covino E., Terranegra L., Vittone A., and Vrba F.  (1986) {\it Astrophys. J., 310}, L71-L75.

\refs Herbst W., Booth J. F., Koret D. L., Zajtseva G. V.  et al. (1987)  {\it Astron. J.,  94}, 137-149.

\refs Herbst W.,  Herbst W., Grossman E. J., and Weinstein D. (1994) 
{\it Astron. J., 108}, 1906-1923.
  
\refs Herbst W., Rhode K. L., Hillenbrand L. A., and Curran G. (2000a)
{\it Astron. J., 119}, 261-280.

\refs
Herbst W., Maley J. A., and Williams E. C. (2000b)
 {\it Astron. J., 120}, 349-366.

\refs Herbst W., Bailer-Jones C. A. L., and Mundt R. (2001) 
 {\it Astrophys. J., 554}, L197-L200.

\refs Herbst W., Bailer-Jones C. A. L., Mundt R., Meisenheimer K., and
Wackermann R. (2002) {\it Astron. Astrophys.,  396}, 513-532.

\refs Herbst W.,  Francis A., Dhital S.,Tresser N., Lin L, and Williams  
E. C. (2005) {\it Bull. Am. Astron. Soc., 37}, 74.10.

 \refs Joergens V., Fern\'andez M., Carpenter J. M., and Neuh{\"a}user R. (2003)
{\it Astrophys. J., 594}, 971-981.

 \refs Johns-Krull. C. M. and Gafford A. (2002)
{\it Astrophys. J., 573}, 685-698.

\refs Kirkpatrick J. D., Reid I. N., Liebert J., Cutri R. M., Nelson B., 
Beichman C. A., Dann C. C., Monet D. G., Gizis J. E., and Skrutskie M. F. (1999)
{\it Astrophys. J., 519}, 802-833.

\refs Kiziloglu \"U, Kiziloglu N., and Baykal A (2005) 
{\it Astron. J., 130}, 2766-2777.

\refs K\"onigl A. (1991) {\it Astrophys. J., 370}, L39-L43.

\refs Krishnamurthi A., Pinsonneault M. H., Barnes S., and Sofa S.  (1997)
{\it Astrophys. J., 480}, 303-323.

\refs Lamm M. H. (2003), {\it Ph.D. Thesis}, University of Heidelberg

\refs Lamm M. H., Bailer-Jones C. A. L., Mundt R., Herbst W., 
and Scholz A. (2004)  {\it Astron. Astrophys., 417}, 557-581.

\refs Lamm M. H., Mundt R., Bailer-Jones C. A. L., and  Herbst W. (2005)
 {\it Astron. Astrophys., 430}, 1005- 1026. 
 
 \refs Littlefair S. P., Naylor T., Burningham B., and Jeffries R. D. (2005)  {\it Mon. Not. R. Astron. Soc., 358}, 341-352.
 
 \refs Long M., Romanova M. M., and Lovelace R. V. E. (2005) {\it Astrophys. J., 634}, 1214-1222.
 
 \refs L\'opez Mart{\'\i} B., Eisl{\"o}ffel J., Scholz A., and Mundt R. (2004) 
{\it Astron. Astrophys., 416}, 555-576.

\refs 
Makidon R. B., Rebull L. M., Strom S. E., Adams M. T., and 
Patten B.M. (2004)  {\it Astron.. J., 127}, 2228-2245. 

\refs Mandel G. N. and Herbst W. (1991) {\it Astrophys. J. 383}, L75-78.

\refs Mart\'{\i}n E. L. and Zapatero-Osorio M. R. (1997)
{\it Mon. Not. R. Astron. Soc., 286}, 17-20.

\refs Mathieu R. D.  (2004) in
{\it Stellar Rotation, Proceedings of IAU Symposium No. 215}, pp. 113-122, Aston. Soc. Pacific), San Francisco.

\refs Matt S. and Pudritz R. E. (2004) 
{\it Astrophys. J. 632}, L135-138. 

\refs Matt S. and Pudritz R. E. (2005a) 
{\it Mon. Not. R. Astron. Soc., 356}, 167-182. 

\refs Matt S. and Pudritz R. E. (2005b) {\it Astrophys. J., 632}, L135-138. 

\refs Mohanty S. and Basri G. (2003) {\it Astrophys. J., 583}, 451-472.

\refs Mohanty S., Jayawardhana R., and Basri G. (2005a)
{\it Mem. Soc. Astron. Ital., 76}, 303-308.

\refs Mohanty S., Jayawardhana R., and Basri G. (2005b) {\it Astrophys. J., 626}, 498-522.

\refs Mundt R. (1984)  {\it Astrophys. J., 280}, 749-770.

\refs Nakajima T., Oppenheimer B. R., Kulkarni S. R., Golimowsk, D. A., 
Matthews K., and Durrance S. T. (1995) 
{\it Nature, 378}, 463.

\refs Phan-Bao N., Guibert J., Crifo F., Delfosse X., Forveille T., 
Borsenberger J., Epchtein N., Fouqu\'e, and P. Simon, G. (2001)
{\it Astron. Astrophys., 380}, 590-598. 

\refs Rebolo R., Zapatero-Osorio M. R. and Mart\'{\i}n E. L. (1995)
{\it Nature, 377}, 129.

\refs 
Rebull L. M. (2001) 
{\it Astron. J., 121}, 1676-1709. 

\refs 
Rebull L. M., Wolff S. C., Strom S. E., and Makidon R. B.  (2002) 
{\it Astron. J., 124}, 546-559. 

\refs 
Rebull L. M., Wolff S. C., and Strom S. E.  (2004) 
{\it Astron. J., 127}, 1029-1051. 

\refs 
Rebull L. M., Stauffer, J. R., Megeath, T., Hora, J., and Hartmann, L. (2005) 
{\it Bull. Am. Astron. Soc. 207}, 185.08.

\refs 
Rhode K. L., Herbst W., and Mathieu R. D. (2001) {\it Astron. J., 122}, 3258-3279. 

\refs 
Rydgren A. E.  and Vrba F. J.  (1983) 
{\it Astrophys. J., 267}, 191-198. 

\refs Scholz A. (2004)
{\it Ph.D. thesis}, University of Jena.

\refs Scholz A. and Eisl{\"o}ffel J. (2004a)
{\it Astron. Astrophys., 419}, 249-267.

\refs Scholz, A. and Eisl{\"o}ffel J. (2004b)
{\it Astron. Astrophys., 421}, 259-271.

\refs Scholz A. and Eisl{\"o}ffel J. (2005)
{\it Astron. Astrophys., 429}, 1007-1023.

\refs Sherry W. H. (2003)
{\it Ph.D. thesis}, State University of New York at Stony Brook.

\refs  
Shu F., Najita J., Ostriker E., Wilkin F., Ruden S., and Lizano S. 
(1994)  {\it Astrophys. J., 429}, 781-796. 

\refs
Sills A., Pinsonneault M. H., and Terndrup D. M. (2000)  {\it Astrophys. J., 543}, 335-347.

 \refs Stassun K. G. and Terndrup D.(2003) 
{\it Pub. Astron. Soc. Pac., 115}, 505-512. 
 
 \refs Stassun K. G., Mathieu R. D., Mazeh T., and Vrba F.J.(1999) {\it Astron. J., 117}, 2941-2979. 

 \refs Stassun K. G., Mathieu R. D., Vrba F.J., Mazeh T., and Henden A. (2001) 
{\it Astron. J., 121}, 1003-1012. 

\refs  
Steinhauer A. J. B., Herbst W., and  Henden A.(1996) {\it Bull. Am. Astron. Soc., 28}, p.884.

\refs 
Strassmeier K. G. (1992) in {\it Robotic telescopes in the 1990s}, pp. 39-52, Astron. Soc. of the Pacific, San Francisco. 

\refs  
Tackett S., Herbst W., and Williams E. C. (2003) {\it Astron. J., 126}, 348-352.

 \refs Tassoul J.-L. (2000) in
{\it Stellar Rotation}, Cambridge U. Press, New York.

\refs Terndrup D. M., Krishnamurthi A., Pinsonneault M. H., and
Stauffer J. R. (1999)
{\it Ann. Rev. Astron. Astrophys., 41}, 599-643.
 
 \refs Terndrup D. M., Stauffer J. R., Pinsonneault M. H., Sills A., 
Yuan Y., Jones B. F., Fischer D., and Krishnamurthi A. (2000)
{\it Astron. J., 119}, 1303-1316.

\refs Thompson M. J., Christensen-Dalsgaard J., Miesch M. S., and Toomre J. (2003)
{\it Astron. J., 118}, 1814-1818.

\refs Tinker J., Pinsonneault M., and Terndrup, D. M. (2002)
{\it Astrophys. J., 564}, 877-886.

\refs  Uzdensky D.A. (2004)  {\it Astrophys. and Space Sci.,  292}, 573-585. 

\refs Vogel S. N. and Kuhi L. V. (1981) {\it Astrophys. J., 245}, 960-976.

\refs Vrba, F. J., Herbst W., and Booth J. F. (1988) 
{\it Astron. J., 96}, 1032-1039.

\refs Whelan E. T., Ray T. P., Bacciotti F., Natta A., Testi L., and Randich S.
      2005  {\it Nature, 435},  652-654.

\refs Woitas J., Bacciotti F., Ray T. P., Marconi A., Coffey D., 
      and Eisl\"offel J. (2005)  {\it Astron. Astrophys., 432}, 149-160.
      
\refs Zapatero-Osorio M. R., Rebolo R., and Mart{\'\i}n E. L. (1997
{\it Astron. Astrophys., 317}, 164-170.

\refs Zapatero-Osorio M. R., Caballero J. A., B{\'e}jar V. J. S., and Rebolo R. 
(2003) {\it Astron. Astrophys., 408}, 663-673.

\refs  
}

\end{document}